\documentclass[pageno]{jpaper}

\usepackage{cite}

\usepackage{amssymb,amsmath,amsfonts}
\usepackage{comment}
\usepackage{graphicx}
\usepackage{subfig}
\usepackage{ctable}
\usepackage{color}
\usepackage{soul}

\usepackage{algorithmicx}
\usepackage{algorithm}
\usepackage{algpseudocode}

\newcommand{\shrink}{Eliminate vertical white-space}
\newcommand{\vshrink}[1]{
  \ifdefined\shrink
    \vspace{-#1cm}
  \else
    \vspace{0cm}
  \fi
}

\newcommand{\POWTEN}[1]{10\textsuperscript{#1}}
\newcommand{\arch}{AISC}

\newcommand{\x}{{\em Breadth}}
\newcommand{\y}{{\em Depth}}

\title{AISC: Approximate Instruction Set Computer}

\author{
  Alexandra Ferrer\'on$^1$, Jes\'us Alastruey-Bened\'e$^1$, Dar\'io Su\'arez-Gracia$^1$, Ulya R. Karpuzcu$^2$\\
  $^1$ Universidad de Zaragoza, Spain~~~~~~~~~$^2$ University of Minnesota, Twin Cities
}

\begin{document}

\date{}
\maketitle

\thispagestyle{plain}
\pagestyle{plain}

\noindent \begin{abstract}
 \noindent This paper makes the case for a single-ISA heterogeneous computing platform,
\arch, where each compute engine (be it a core or an accelerator) supports a
{different subset} of the {very same} ISA.  An ISA subset may not be
functionally complete, but the union of the (per compute engine) subsets renders
a functionally complete, platform-wide {\em single} ISA. Tailoring the
microarchitecture of each compute engine to the subset of the ISA that it
supports can easily reduce hardware complexity.  At the same time, the energy
efficiency of computing can improve by exploiting algorithmic noise tolerance:
by mapping code sequences that can tolerate (any potential inaccuracy induced
by) the incomplete ISA-subsets to the corresponding compute engines.

\end{abstract}

\section{Introduction}
  \label{sec:introduction}
  The ISA specifies semantic and syntactic characteristics of a {\em practically}
functionally complete set of machine instructions.
The ISA specifies semantic and syntactic characteristics of a {\em practically}
{functionally} complete set of machine instructions. Modern ISAs are not
necessarily {\em mathematically} functionally complete, but provide sufficient
expressiveness for practical algorithms. For software layers, the ISA defines
the underlying machine -- as capable as the variety of algorithmic tasks the
composition of its building blocks, {\em instructions}, can express. For hardware
layers, the ISA rather acts as a behavioral design specification for the machine
organization. Accordingly, the ISA governs both the functional completeness and
complexity of a machine design.

This paper makes the case for an alternative, single-ISA heterogeneous computing
platform, \arch, 
which can reduce the ISA complexity, and thereby improve energy efficiency, on a per
compute engine (be it a core or an accelerator) basis, without compromising the
functional completeness of the overall platform.  The distintinctive feature of
\arch\ 
is that each compute engine supports a {\em  different subset} of the {\em very
same} instruction set. Such per compute engine ISA subsets may be disjoint or
overlapping. An ISA subset may not be functionally complete, but the union of
the (per compute engine) subsets renders platform-wide a functionally complete
{\em single} ISA. Therefore, software layers perceive \arch\ as a single-ISA
machine.  
Tailoring the microarchitecture of each compute
engine to the subset of the ISA that it supports results in less complex,
more energy efficient compute engines, without compromising the overall
functional completeness of the machine.  

When it comes to the design of a feasible AISC platform, many questions 
arise, the most critical being:
\begin{list}{\labelitemi}{\leftmargin=0.5em}
\item Which subset of the ISA should each compute engine support, by
  construction?
  \item How to guarantee that each sequence of instructions scheduled to execute
	on a given compute engine only spans the respective subset of the ISA (with
	potential accuracy loss)?
  More specifically, how to map instruction sequences to the compute engines?
  \item How to orchestrate migration of code sequences from one
	compute engine to another within the course of computation, as different application
	phases may exhibit different degrees of tolerance to noise?
  \item	How to keep the potential accuracy loss bounded? 
\end{list}
\vspace{-0.1cm}

Starting with the first and most basic question, approximation along
two dimensions can set the ISA subset per compute engine: 
\vspace{-0.1cm}
\begin{list}{\labelitemi}{\leftmargin=0.5em}
\item  \x\ approximation simplifies instructions by reducing
  complexity (e.g., precision) on a per instruction basis. To be more specific,
  the subset of the ISA a compute engine implements in this case would selectively contain
  lower complexity (e.g., lower precision) instructions, by construction.
Well-studied precision reduction approaches
~\cite{Yeh2009, Chippa2014, Sampson2011,
Tong1998, Rubio-Gonzalez2013, Moreau2016, Jain2016, Stephenson2000}
are directly applicable in this context.
Reducing the operand width often enables simplification in the
corresponding arithmetic operation, in addition to a more efficient utilization
of the available communication bandwidth for data (i.e., operand)
transfer. 

\item  \y\ approximation excludes complex and less frequently used
  instructions from the subset.

\item The combination of the two dimensions, \x\ + \y\ , is
also possible: In this case, the compute engine concerned would be able to {\em
approximately} emulate complex and less frequently used instructions (that its
ISA subset does not contain) by a sequence of simpler instructions. 
\end{list}
\vspace{-0.1cm}

AISC trades computation accuracy for the complexity (and thereby,
energy efficiency) on a per compute engine basis.  
At the same time, as the entire platform still supports the full-fledged ISA,
instruction sequences not prone to approximation can still run at full accuracy. 
AISC can also be regarded as an aggressive variant of
architectural core salvaging~\cite{acs} or ultra-reduced instruction set
coprocessors~\cite{urisc}, where actual hardware faults impair a compute
engine's capability to implement a subset of its ISA (and all compute engines
support the same ISA by design). These lines of studies detail how to
achieve full-fledged functional completeness under the hardware-fault-induced
loss of support for a subset of instructions. AISC, on the other hand, features
compute engines with approximate, i.e., incomplete or restricted, ISAs by construction.
Without loss of generality, such incomplete or restricted ISAs within an AISC
platform may be due to errors or simply enforced by design. The latter applies
for the following discussion. 

In the following, Section~\ref{sec:impl} details a proof-of-concept AISC
implementation and practical limitations; Sections~\ref{sec:setup} and
~\ref{sec:eval} quantify the complexity vs.\ energy efficiency trade-off as
induced by AISC;
Section~\ref{sec:rel} compares and contrasts AISC with related work; and
Section~\ref{sec:conc} concludes the paper by summarizing our findings.

  \section{Proof-of-Concept Implementation}
  \label{sec:impl}
\begin{figure}[htp]
  \begin{center}
    \subfloat[Native]{\label{fig:hor:baseline}
      \includegraphics[width=0.10\textwidth]{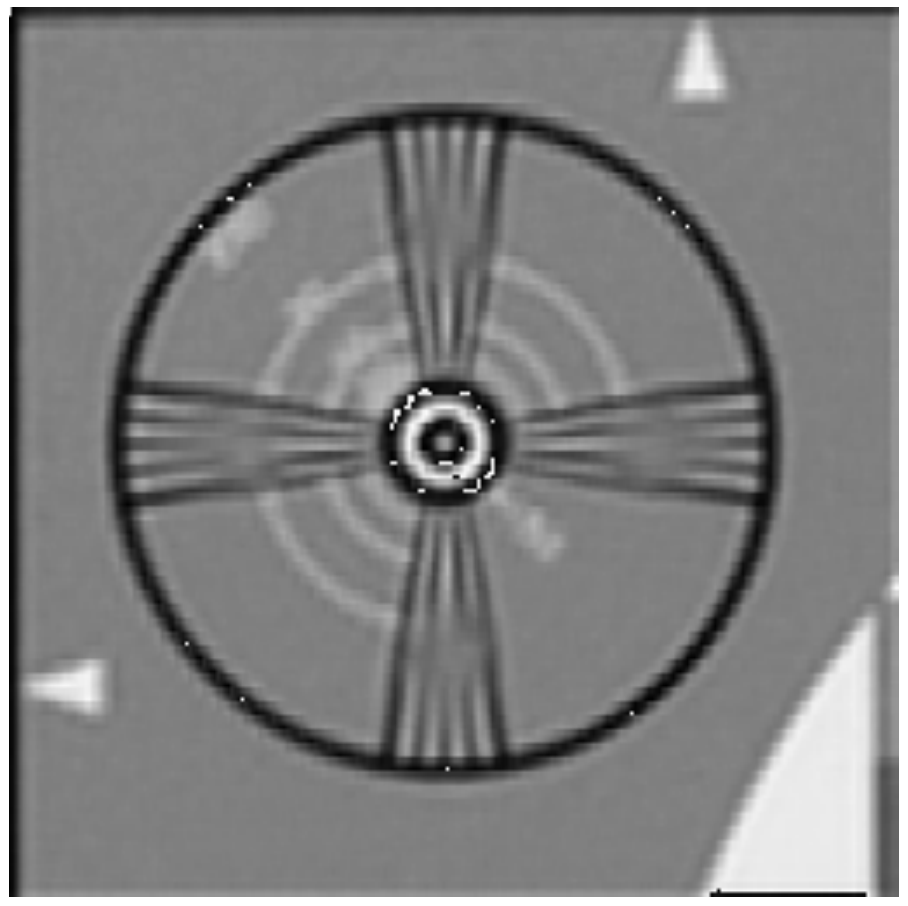}
    } \hfill
	\subfloat[{\em \y}]{\label{fig:hor:drop}
        \includegraphics[width=0.10\textwidth]{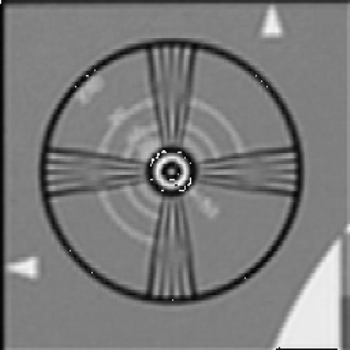}
    } \hfill
	\subfloat[{\em \x}]{\label{fig:hor:dptohp}
      \includegraphics[width=0.10\textwidth]{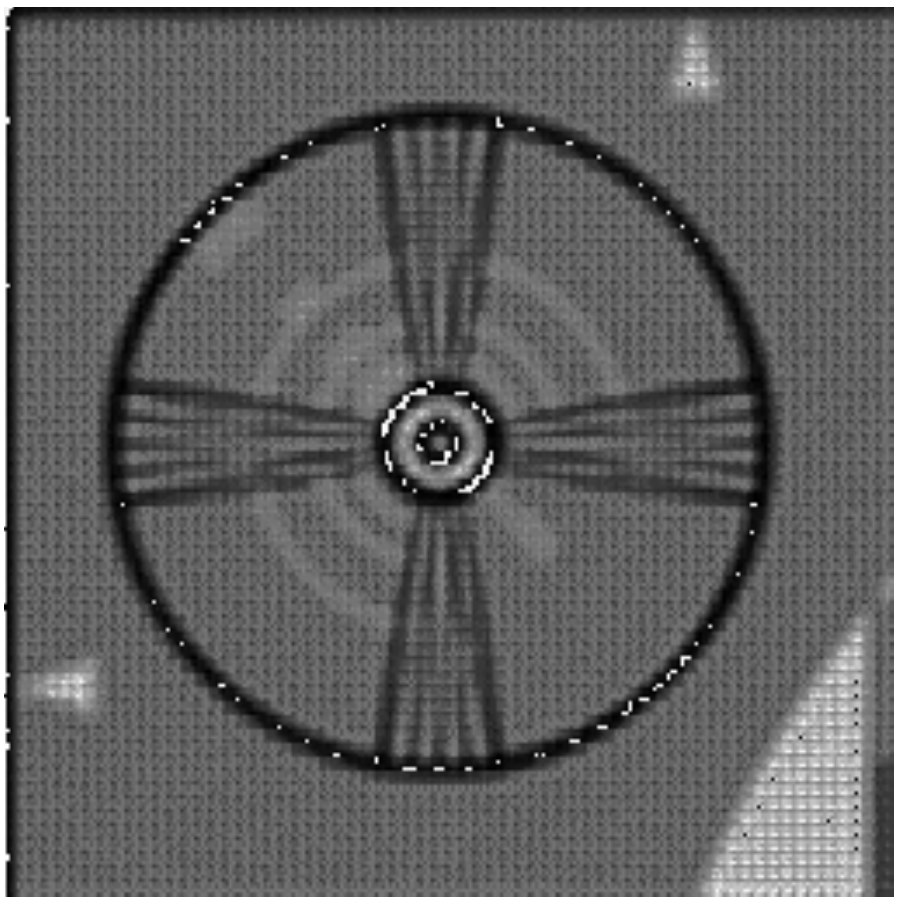}
    } \hfill
	\subfloat[{\em \x}+{\em \y}]{\label{fig:hor:divtomul}
      \includegraphics[width=0.10\textwidth]{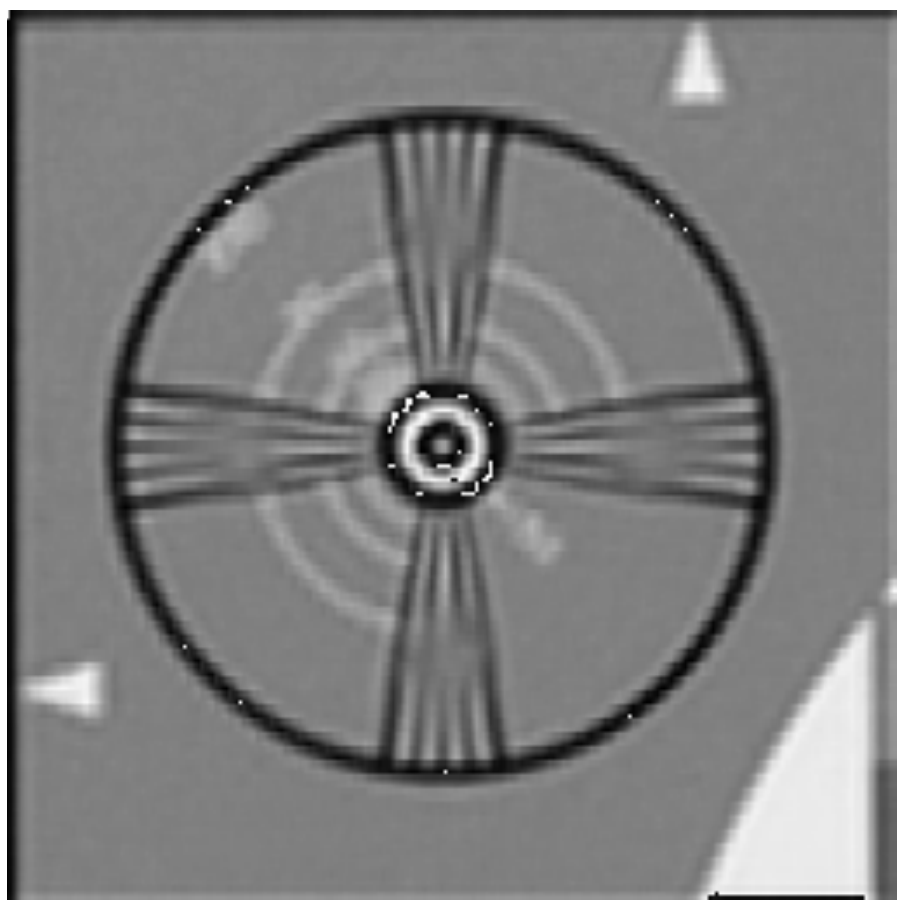}
    }
	\vshrink{0.1}
	\caption{Graphic output of SRR benchmark (Table~\ref{tab:apps}) under
	  representative AISC techniques (b)-(d).
    \label{fig:hor}}
  \end{center}
\end{figure}

  \noindent Let us start with a motivating example.  Fig.~\ref{fig:hor} shows how
the (graphic) output of a typical noise-tolerant application,
SRR (see Table~\ref{tab:apps}), changes for representative \y,
\x, and \x\ + \y\ techniques.
Section~\ref{sec:setup} provides experimental
details.  Fig.~\ref{fig:hor:baseline} captures the output for the baseline for
comparison, {\em Native}\/ execution, which excludes any approximation.
The accuracy loss remains
barely visible but still varies across different techniques.
Let us next take a closer look at the sources of this diversity.

\vshrink{0.4}
\vshrink{0.5}

\subsection{Depth Techniques}

\noindent \y\ encompasses all techniques that orchestrate dropping of complex and
less frequently used instructions.  The key question is how to pick the
instructions to drop.  A more general version of this question, \emph{which
instructions to approximate under AISC}, already applies to AISC techniques
across all dimensions,
but the question becomes more critical for \y.  As the most aggressive in
our bag of tricks, \y\ can incur significant loss in accuracy.  The targeted
recognition-mining-synthesis (RMS) applications
can tolerate errors in data-centric phases as opposed to control~\cite{Cho2012}.
Therefore, confining instruction dropping to data-flow can help limit the
incurred accuracy loss.  

\subsection{Breadth Techniques}
\noindent Without loss of generality, we experimented with three approximations
to reduce operand widths: \emph{DPtoSP},  \emph{DP(SP)toHP}, and \emph{DP(SP)toINT}.

Under the IEEE 754 standard, 32 (64) bits express a single
(double) precision floating point number: one bit specifies the \emph{sign}; 8
(11) bits, the \emph{exponent}; and 23 (52) bits the \emph{mantissa}, i.e., the
fraction.  For example, $(-1)^{sign}\times2^{exponent-127}\times1.mantissa$
represents a single-precision floating number.
DPtoSP
is a {\em bit discarding} variant, which
omits 32 least-significant bits of
the mantissa of each double-precision operand of an instruction, and keeps the exponent
intact.
DP(SP)toHP comes in two flavors. DPtoHP
omits 48 least-significant bits of
the mantissa of each double-precision operand of an instruction, and keeps the exponent
intact;  SPtoHP, 16 least-significant bits of
the mantissa of each single-precision operand of an instruction.
Fig.~\ref{fig:hor:dptohp} captures an example execution outcome under
DPtoHP.
DP(SP)toINT also comes in two flavors. DPtoINT (SPtoINT)
replaces double (single) precision instructions with their integer counterparts,
by rounding the floating point operand values to the closest integer.

\begin{algorithm}[H]
\caption{\textbf{DIVtoMUL.NR}}\label{alg:nr}
\footnotesize
; Take reciprocal of the divisor: \\
; $x_0 = RCP(divisor)$ ; 12-bit precision \\
; Newton-Raphson iteration to increase reciprocal precision\\
; to 23 bits: \\
; $x_1 = x_0 \times (2-divisor \times x_0) = 2 \times x_0 - divisor \times x_0^2$ \\
; Multiply dividend with reciprocal: \\
; $result = dividend \times x_1$ \\
1 \textbf{MOVSS xmm1, divisor} \\
2 \textbf{RCPSS ~xmm0, xmm1}~~~~~~~~~~; $x_0$ \\
3 \hl\textbf{MULSS xmm1, xmm0}~~~~~~~~~~~; $divisor \times x_0$ \\
4 \hl\textbf{MULSS xmm1, xmm0}~~~~~~~~~~~; $(divisor \times x_0) \times x_0$ \\
5 \hl\textbf{ADDSS xmm0, xmm0}~~~~~~~~~~~; $2 \times x_0$ \\
6 \hl\textbf{SUBSS ~xmm0, xmm1}~~~~~~~~~~~; $x_1 = 2 \times x_0 - divisor \times x_0^2$ \\
7 \textbf{MULSS xmm0, dividend}~~~~~; $result = dividend \times x_1$
\end{algorithm}
\vspace{-.3cm}

\subsection{Breadth + Depth Techniques}

\noindent The proof-of-concept AISC design features two representative \x\ + \y\ techniques:
\emph{MULtoADD} and \emph{DIVtoMUL}.
MULtoADD converts multiplication instructions to a sequence of additions.
MULtoADD picks the smaller of the factors as the multiplier, which determines the number of additions.
MULtoADD rounds floating point multipliers to the closest integer number.
DIVtoMUL converts division instructions to multiplications.
DIVtoMUL first calculates the reciprocal of the divisor, which next gets multiplied by the dividend to render the end result.
In our proof-of-concept implementation based on the x86 ISA, the reciprocal instruction has 12-bit precision.
\emph{DIVtoMUL12} uses this instruction.
\emph{DIVtoMUL.NR}, on the other hand, relies on one iteration of the Newton-Raphson method~\cite{Ercegovac2004} to increase the precision of the reciprocal to 23 bits.
DIVtoMUL12 can be regarded as an approximate version of DIVtoMUL.NR, eliminating the Newton-Raphson iteration, and hence enforcing a less accurate estimate of the reciprocal (of only 12 bit precision).
Fig.~\ref{fig:hor:divtomul} captures an example execution outcome under DIVtoMUL.NR.
Algorithm~\ref{alg:nr} provides an example sequence of instructions to emulate division
according to DIVtoMUL.NR, where one reciprocal (\texttt{RCPSS}), 3 multiplication (\texttt{MULSS}), one addition (\texttt{ADDSS}), and one subtraction (\texttt{SUBSS}) instruction substitute a division.
DIVtoMUL12 omits
the iteration of the Newton-Raphson method (lines 3-6) from Algorithm~\ref{alg:nr}.
As opposed to DIVtoMUL.NR, DIVtoMUL12 keeps the 12-bit accuracy of the \texttt{RCPSS} instruction.
Hence, it becomes mathematically equivalent to omitting 11 bits of the mantissa.

  \section{Evaluation Setup}
  \label{sec:setup}
  \setlength\tabcolsep{2.5pt} 

\ctable[
    caption = {Benchmarks deployed.}, 
    mincapwidth = \textwidth,
    label = tab:apps,
    maxwidth = \textwidth,
    doinside = \scriptsize,
    pos = !htbp,
    footerwidth = .95\textwidth,
    star
]
{@{}lm{2.1cm}m{6.0cm}m{4.0cm}@{}}
{
}
{
    \toprule
    \textbf{Application} & \textbf{Domain} & \textbf{Input} & \textbf{Output}\\
    \midrule
    K-means (KM) & Clustering & Edge features Corel Corporation DB (100 entries)
        & Cluster assignments\\
    Latent Dirichlet Allocation (LDA) & Topic Modeling 
        & 500 documents, 6097 terms; variat. inference: \newline 20 itr./\POWTEN{-6} error; variat. EM: 100 itr./\POWTEN{-5} error
        & Estimation Model \\
    Motion-Estimation (ME) & Computer Vision & ``Alpaca" Dataset (16 frames, 96 x 128)
        & Motion vectors (1 per frame)\\
    Principal Component Analysis (PCA) & Feature Selection & Handwritten digit (1593 instances, 256 ATR)
        &  Column- \& Row-Projections\\
    Restricted Boltzmann Machine (RBM)  & Deep Learning 
        & Netflix DB (100 train users x 100 movies \newline x 20 loops; 100 test users x 100 movies)
        & Suggestions for \newline users/movies (100 x 100 matrix) \\
    Super-Resolution Reconstruction (SRR) & Computer Vision & ``EIA" Dataset (16 frames, 64 x 64)
        & Reconstructed Image (256 x 256)\\
    Single Value Decomposition (SVD3) & Feature Selection & KOS Press (500x500 matrix)
        & Decomposition matrices \\ \bottomrule
}

\noindent We analyze a representative mix of RMS benchmarks from Cortex Suite~\cite{Thomas2014}, as tabulated in Table~\ref{tab:apps}, compiled with GCC 4.8.4 with \texttt{-O1\@} on an Intel\textsuperscript{\textregistered} Core\texttrademark~i5 3210M machine.
As we perform manual transformations on the code, high optimization levels hinder the task; we resort to \texttt{-O1} for our proof-of-concept and leave for future work more exploration on compiler optimizations.
For each application, we focus on the main kernels (i.e., region of interest) where the actual computation takes place.
Throughout the evaluation, we map the region of interest of an application in its entirety
to an incomplete-ISA (AISC) compute engine, irrespective of potential changes in noise
tolerance within the course of its execution.
We use ACCURAX metrics~\cite{Akturk2015} to quantify the accuracy-loss.
We prototype AISC techniques from Section~\ref{sec:impl} on Pin
2.14~\cite{Luk2005}, which is based on the x86 ISA.

Our energy model follows~\cite{Czechowski2014}, which is based on bare-metal measurements for x86 processors.
The model distinguishes between fixed and variable components of energy consumption.
The variable component captures changes in activity.
Beyond activity, the ratio of fixed to variable components depends on technology and microarchitecture.
For example, on Intel Penryl, 56\% of the energy consumption is due to the fixed component, while on Haswell this percentage increases to 75\%~\cite{Czechowski2014}.
AISC techniques can affect both, the fixed and the variable component.
Savings in the variable component come from the operand width reduction under \x, and from the instruction dropping under \y.
The fixed component can also decrease if the microarchitecture exploits AISC for a less complex pipeline front-end (fetch + decode) design.
In the evaluation, we will only report anticipated energy savings in the variable component.
The overall impact of the variable component on energy savings depends on the ratio of the fixed and variable components.
We represent energy in terms of number of instructions $\times$ EPI (energy per instruction).
We conservatively assume that the EPI of an integer instruction equals the EPI of a 32-bit floating point arithmetic instruction, and scale the EPI values for 64-bit and 16-bit operations according to~\cite{Choi2013}.

  \section{Evaluation}
  \label{sec:eval}
  \begin{figure*}[!ht]
\centering
\includegraphics{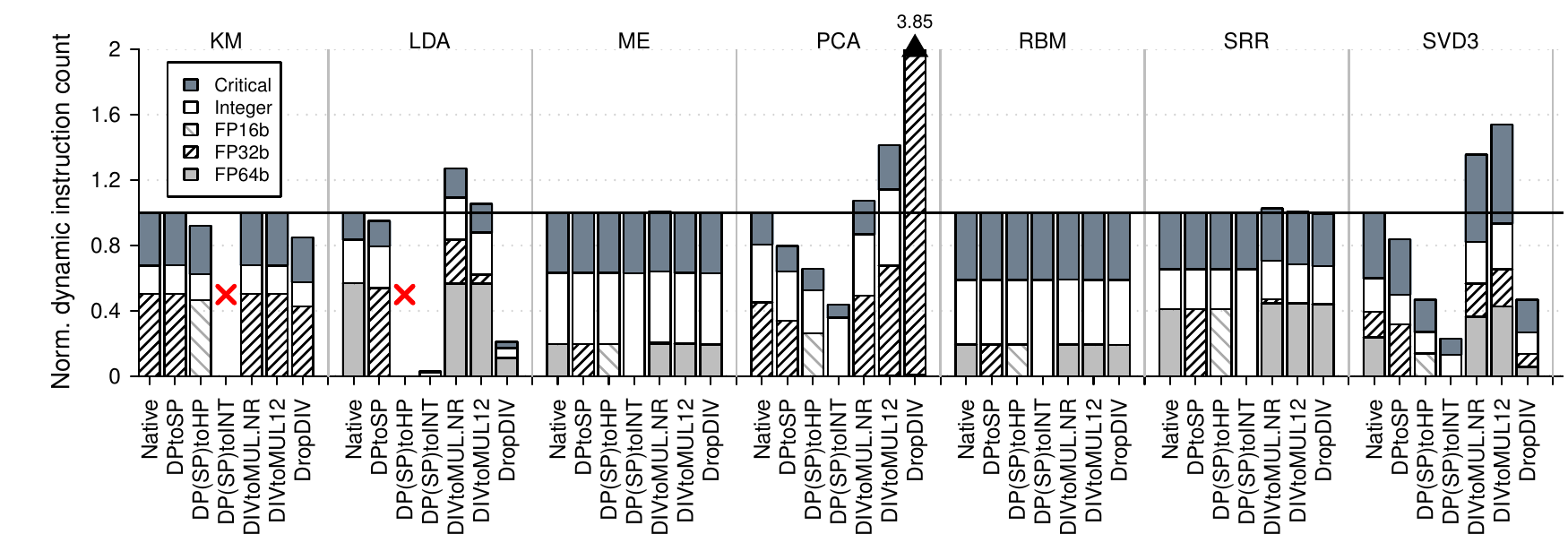}
\vshrink{0.4}
\caption{Impact on instruction mix and count.}
\label{fig:imix}
\end{figure*}

\begin{figure*}[!ht]
\centering
\subfloat[KM]{\label{fig:ener:km} \includegraphics[width=.21\textwidth]{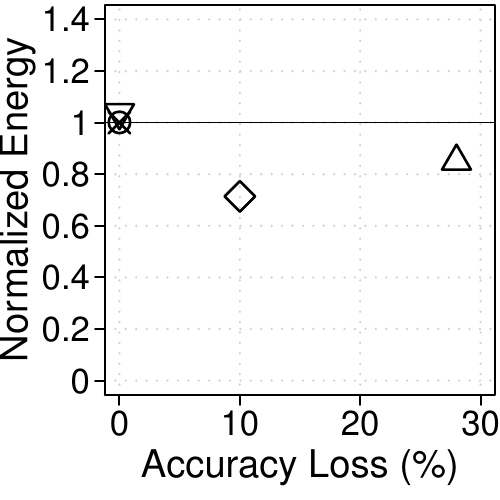}}
~~\subfloat[LDA]{\label{fig:ener:lda} \includegraphics[width=.21\textwidth]{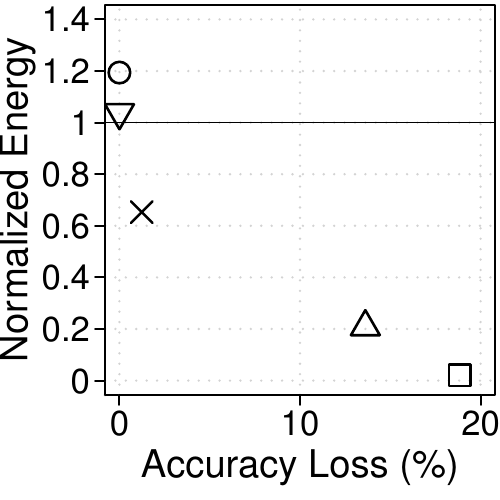}}
~~\subfloat[ME]{\label{fig:ener:me}\includegraphics[width=.21\textwidth]{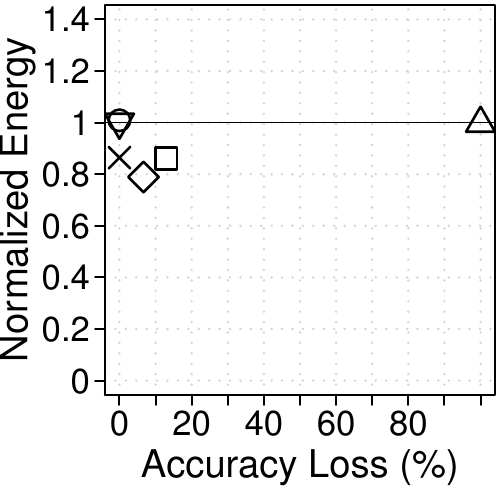}} \\ 
\subfloat[PCA]{\label{fig:ener:pca}\includegraphics[width=.21\textwidth]{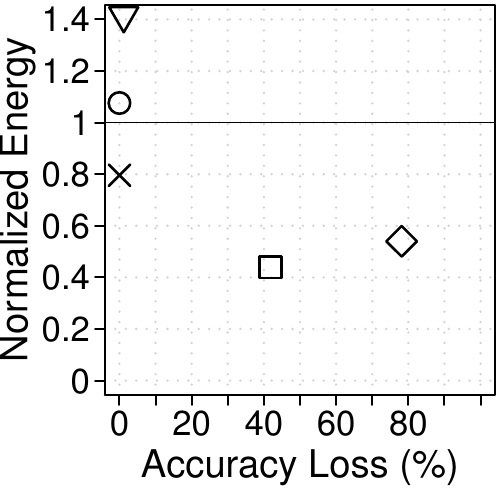}} 
~~\subfloat[RBM]{\label{fig:ener:rbm} \includegraphics[width=.21\textwidth]{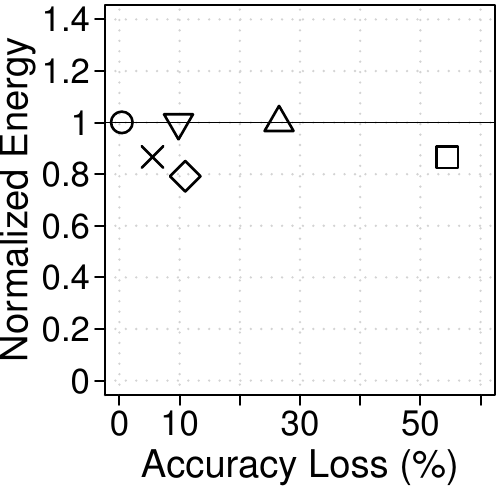}}
~~\subfloat[SRR]{\label{fig:ener:srr}\includegraphics[width=.21\textwidth]{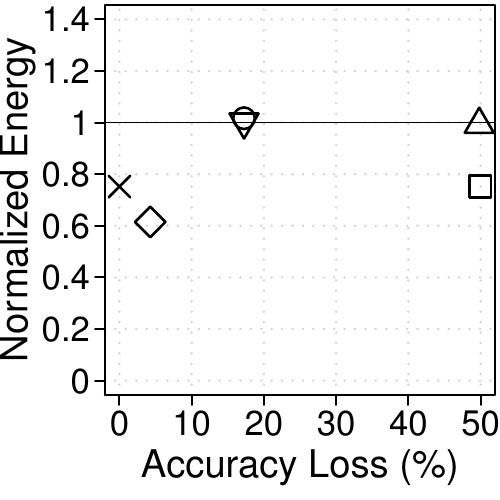}} \\
\subfloat[SVD3]{\label{fig:ener:svd}\includegraphics[width=.21\textwidth]{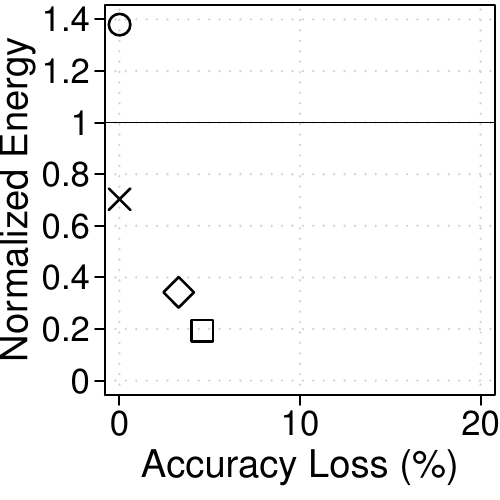}} 
\subfloat[Legend]{\label{fig:ener:leg} \includegraphics[width=.41\textwidth]{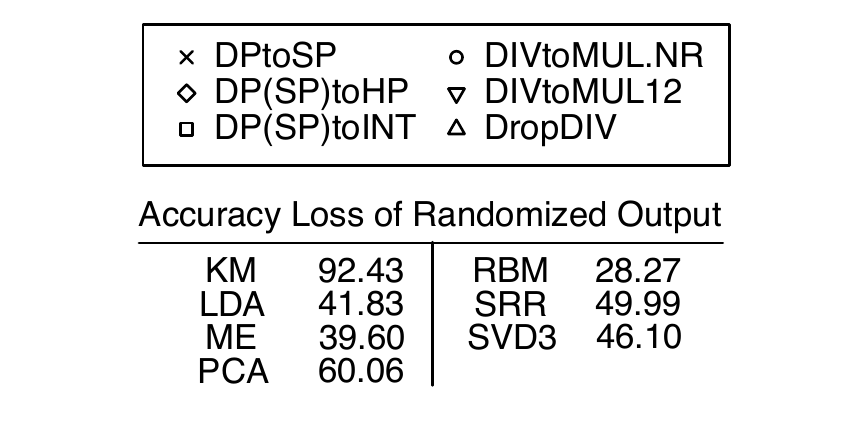}}
\caption{Energy vs. Accuracy trade-off.}
\label{fig:ener}
\end{figure*}

\noindent Fig.~\ref{fig:imix} shows the impact on instruction mix and count, as characterized by a group of bars for each benchmark.
The first bar corresponds to the \emph{Native} execution outcome, excluding any AISC technique.
The rest of the bars capture execution under \x, \x\ + \y, and \y.
The height of each bar captures the relative change in the instruction count with respect to the native outcome.
The stacks in each bar depict the instruction mix, considering three categories:
\emph{Critical} indicates control-flow instructions such as accesses to the
stack;
\emph{Integer}, integer data transfer and arithmetic;
\emph{FP (Floating Point)}, floating point data transfer and arithmetic.
AISC excludes \emph{Critical} instructions, as approximating or removing them might alter the control flow and prevent successful program termination.
The proof-of-concept AISC implementation does not approximate \emph{Integer} instructions to avoid corrupting pointer arithmetic, which may in turn prevent successful termination.
We further categorize \emph{FP} instructions by the size of the operands: 64b(it), 32b, and 16b.

To understand the corresponding implications for accuracy, Fig.~\ref{fig:ener}
shows, on a per benchmark basis, the trade-off space of energy versus accuracy loss.
Each point corresponds to the outcome under one AISC technique.
We report both energy and accuracy loss relative to the native execution outcome, which excludes AISC techniques.
Trade-off spaces do not include non-terminating executions and executions that render more than 40\% increase in energy consumption.
On the normalized energy axis, any point above 1 is unfeasible.
Fig.~\ref{fig:ener:leg} shows the output randomization results for each benchmark~\cite{Akturk2015}, which serve as a proxy for (close to) worst case accuracy loss.

Next, we examine the proof-of-concept AISC techniques from
Section~\ref{sec:impl}
in more detail.

\subsection{Breadth Techniques}

\noindent We observe that \x\, in terms of DPtoSP, DP(SP)toHP, and DP(SP)toINT, can reduce the instruction count significantly for
benchmarks PCA and SVD3 (Fig.~\ref{fig:imix}).
These benchmarks adapt iterative refinement; under bit discarding due to \x, they reach the convergence criteria earlier.
In this case, only two experiments fail to terminate, marked by a cross in Fig.~\ref{fig:imix}: DP(SP)toINT in KM and DP(SP)toHP in LDA.
Significant changes in instruction count mainly stem from early or late convergence.

\x, in terms of DPtoSP and DP(SP)toHP, provides large energy reductions (20\%
and 37\% on average) accompanied by a modest accuracy loss (less than 10\%) for
most of the applications -- except PCA and SRR, where accuracy loss reaches
78.1\% and 34.6\%, respectively; and LDA where the experiments failed to
terminate due to bit discarding.
Even the very aggressive DP(SP)toINT works for LDA, ME, and SVD3, where the accuracy loss becomes 18.9\%, 13.0\%, and 4.6\%, respectively.
The energy reduction for LDA (98\%) and SVD3 (81\%) is significant, as the number of iterations to convergence gets drastically reduced: LDA reduces the number of iterations by 97.3\%; SVD3, by 97.9\%.
KM is the only application that does not survive DP(SP)toINT.
For the rest of the benchmarks, the accuracy loss becomes comparable to the
accuracy of a randomly generated output (which captures close-to-worst-case
accuracy loss~\cite{Akturk2015}), therefore, likely not
acceptable (Fig.~\ref{fig:ener:leg}).

\subsection{Depth Techniques}

\noindent \y\
comes in two flavors.
First, we selectively remove all the floating point division instructions of the binary (\emph{DropDIV}).
This would be equivalent to removing the floating point division instruction
from the ISA, without providing any substitute (as opposed to \x\ + \y~techniques).
Dropping division instructions does not affect the termination of the experiments, although PCA and SVD3 do not reach convergence.
As shown in Fig.~\ref{fig:imix}, KM and LDA show a significant reduction in the executed instructions under DropDIV, which translates into an energy reduction of 15\% and 90\%, respectively, with an accuracy loss of 28\% and 13.62\%.
For ME, RBM, and SRR, the instruction count remains practically the same, while
the accuracy loss of the outputs reaches the loss of completely random outputs
(as indicated Fig.~\ref{fig:ener:leg}).

\begin{figure}[!t]
\centering
\subfloat[Native]{\label{fig:srr:s0} \includegraphics[width=.22\columnwidth]{srrsmallBaseline}}
\subfloat[1 inst.]{\label{fig:srr:s3} \includegraphics[width=.22\columnwidth]{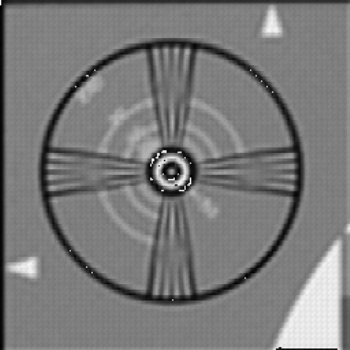}}
\subfloat[3 instr.]{\label{fig:srr:s1} \includegraphics[width=.22\columnwidth]{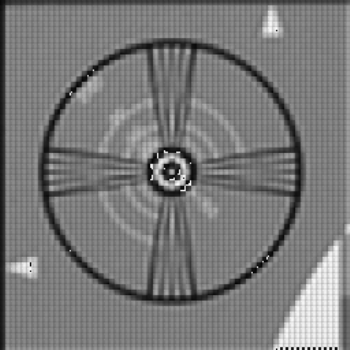}}
\subfloat[5 instr.]{\label{fig:srr:s5} \includegraphics[width=.22\columnwidth]{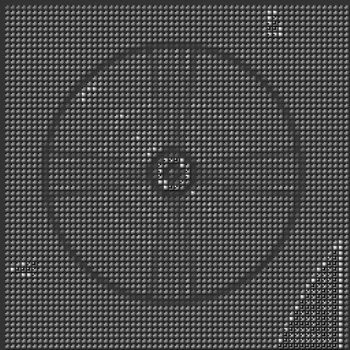}}
\setlength{\belowcaptionskip}{-10pt}
\vshrink{0.2}
\caption{Accuracy of SRR output under \y.}
\label{fig:srr:sd}
\end{figure}

As a more aggressive \y\  approach, we randomly delete static (arithmetic) FP instructions.
For each static instruction, we base the dropping decision on a pre-defined threshold \emph{t}.
We generate a random number \emph{r} in the range $[0, 1]$, and drop the static instruction if \emph{r} remains below \emph{t}.
We experiment with threshold values between 1\% and 10\%.
We observe three distinctive behavior:
\begin{enumerate}
\item SVD3 and PCA do not tolerate \y; experiments either fail to terminate, or render an invalid output/excessive accuracy loss.
\item ME, RBM, and SRR can tolerate dropping, but the outcome highly depends on the instructions dropped.
Fig.~\ref{fig:srr:sd} illustrates three different SRR outcomes for $t$=3\%:
dropping 1 static instruction (Fig.~\ref{fig:srr:s3}); 3 static instructions
(Fig.~\ref{fig:srr:s1}); 5 static instructions (Fig.~\ref{fig:srr:s5}); the native
output is also shown for comparison (Fig.~\ref{fig:srr:s0}).
SRR has 477 static instructions, out of which around 80 are FP.
The dropped static instructions translate into dropping 16 million dynamic instructions
in Fig.~\ref{fig:srr:s3}; 245 million in Fig.~\ref{fig:srr:s1}; and
255 million in Fig.~\ref{fig:srr:s5}, respectively.
The numeric accuracy metric, SSIM~\cite{Akturk2015} becomes 17.1\%
(Fig.~\ref{fig:srr:s3}), 30.2\% (Fig.~\ref{fig:srr:s1}), and 48.2\% (Fig.~\ref{fig:srr:s5}).
To compensate for the missing instructions, SRR executes additional iterations, increasing the dynamic instruction count.
\item For KM and LDA some experiments fail and some survive with varying accuracy loss.
\end{enumerate}

\subsection{Breadth + Depth Techniques}

\noindent Under \x\ + \y, we observe that MULtoADD (Section~\ref{sec:impl}) significantly increases the
instruction count and/or prevents successful termination (LDA and SVD3) -- except for RBM, where most of the multiplications involve very small operands between 0 and 1.
In any case, we did not observe any improvement on the energy vs. accuracy trade-off
space, and, therefore, we excluded MULtoADD from Figs.~\ref{fig:imix} and~\ref{fig:ener}.

DIVtoMUL variants also increase instruction count, although this increase is
only significant for LDA, PCA, and SVD3, where more iterations are run to
compensate for the precision reduction, as we did not alter the convergence criteria.
\x\ + \y\ -- in terms of DIVtoMUL.NR and DIVtoMUL12 -- does not show an energy advantage, mostly due to our conservative energy modeling (we assume that all \emph{FP} instructions of a given precision have the same EPI).
KM, ME, and SRR have a very small percentage of division operations, accordingly, DIVtoMUL variants have minimal impact.
Eliminating the Newton-Raphson iteration under DIVtoMUL.NR only increases the accuracy loss in RBM.
PCA and SVD3 experience a significant energy increase under DIVtoMUL12 (when compared to DIVtoMUL.NR) due to the increasing number of iterations until convergence to meet the convergence criteria.

  \section{Related Work}
  \label{sec:rel}
  Lopes~\emph{et al.} perform a chronological analysis of several x86 applications
and operating systems and show that 30\% of the instructions were rarely used or
become unused over time~\cite{Lopes2015}.  ISA extensions have been proposed in
the context of approximation~\cite{Esmaeilzadeh2012, Kamal2014}.
In~\cite{Esmaeilzadeh2012}, the authors describe an ISA extension to provide
approximate operations and storage.  Kamal~\emph{et al.} extend the ISA with
custom instructions for embedded devices, and allow approximation on those
custom instructions by not meeting timing requirements, as the results might be
still good enough~\cite{Kamal2014}.  Instead of extending the ISA, we explore
the possibilities of reducing the ISA complexity, by reducing the set of
instructions and/or the size of the operands.

Finally, it is worth referring to the work from Schkufza~\emph{et
al.}~\cite{Schkufza2014}.  Their objective is the automatic generation of
different floating point kernels, applying different optimization in an
stochastic way.  In detail, for each computation kernel they apply a series of
transformations that include changes in the opcodes, the operands, swapping
instructions, and even dropping computations.  Contrary to AISC, however, their
goal is to obtain aggressive optimization of high performance applications
running on conventional hardware.

  \section{Conclusion \& Discussion}
  \label{sec:conc}
  \noindent 
The Instruction Set Architecture (ISA) bridges the gap between software and
hardware layers of the system stack.  ISA grows with the addition of new
extensions in \y~(addition of new instructions) and \x~(wider instructions, in
the case of CISC machines) directions.  This growth increases the complexity of
the fetch and decode hardware (front-end), an already power hungry part of the
pipeline, and critical for performance, as it feeds the back-end with
instructions.  Besides, execution units (back-end) and their control policies
also become more complex.

Conventional ISAs are unaware of the intrinsic error
tolerance of many emerging algorithms.  AISC turns around this assumption to
reduce hardware complexity, and, thereby, energy consumption. Our proof-of-concept analysis revealed that,
in its restricted form  -- where the region of interest of an application is mapped in its entirety
to an incomplete-ISA compute engine, irrespective of potential changes in noise
tolerance within the course of its execution -- AISC can cut energy by up to 37\% at around
10\% accuracy loss.

We find that the energy vs. accuracy trade-off is very sensitive to application
convergence characteristics. Therefore, an efficient AISC implementation should
carefully examine convergence criteria.  For example, under \y, applications
that tolerate instruction dropping either compensate for the missing
instructions by executing more to meet the convergence criteria, or exhibit a
very large accuracy loss.

The presented proof-of-concept implementation does not track data dependencies
beyond instruction boundaries.  Operand width reduction under \x\ or \x\ + \y\ is
confined to the input and output operands of the respective instructions.
Transitively adjusting the precision of the variables which depend on or
determine such input and output operands may substantially increase energy
savings.

The most critical design aspect is how instruction sequences should be mapped to
restricted-ISA compute engines, and how such sequences should be migrated
from one engine to another within the course of execution, to track potential
temporal
changes in algorithmic noise tolerance. While fast code
migration is not impossible, if not orchestrated carefully, the energy overhead
of fine-grain migration can easily become prohibitive. 
Therefore, a break-even point in terms of migration frequency and granularity
exists, past which AISC may degrade energy efficiency.

\bibliographystyle{myabbrv}
\bibliography{aisc}

\end{document}